\documentclass[12pt]{article}

\usepackage{graphicx}
\usepackage{times}

\newfam\bboardfam
\font\tenbboard=msbm10  
 \font\sevenbboard=msbm7
   \font\fivebboard=msbm5 
\textfont\bboardfam=\tenbboard
\scriptfont\bboardfam=\sevenbboard
\scriptscriptfont\bboardfam=\fivebboard
\def\bboard{\fam\bboardfam\tenbboard}
\def\R{{\bboard R}}     

\newcommand{\et}{\mbox{$[E_{tot}]$}}
\newcommand{\ft}{\mbox{$[F_{tot}]$}}
\newcommand{\st}{\mbox{$[S_{tot}]$}}
\newcommand{\nM}{\mbox{nM}}
\newcommand{\uM}{\mbox{$\mu$M}}

\newcommand{\ra}{\rightarrow}

\newcommand{\href}{}

\newcommand{\citep}{\cite}

\title{Multi-bit information storage by multisite phosphorylation}

\author{Matthew Thomson,$^{1}$ Jeremy Gunawardena,$^{1\ast}$
\\[0.5em]
\normalsize{$^{1}$Department of Systems Biology, Harvard Medical School}\\
\normalsize{200 Longwood Avenue, Boston, MA 02115, USA.}\\[0.2em]
\normalsize{$\ast$ Corresponding author: {\tt jeremy@hms.harvard.edu}} \\
\normalsize{Tel: {\sf (617) 432 4839}; Fax: {\sf (617) 432 5012}}
}

\date{}

\begin{document}

\maketitle

\begin{abstract}
Cells store information in DNA and in stable programs of gene expression, which thereby implement forms of long-term cellular memory. Cells must also possess short-term forms of information storage, implemented post-translationally, to transduce and interpret external signals. CaMKII, for instance, is thought to implement a one-bit (bistable) short-term memory required for learning at post-synaptic densities. Here we show by mathematical analysis that multisite protein phosphorylation, which is ubiquitous in all eukaryotic signalling pathways, exhibits multistability for which the maximal number of steady states increases with the number of sites. If there are $n$ sites, the maximal information storage capacity is at least $\log_2\lfloor (n+2)/2 \rfloor$ bits. Furthermore, when substrate is in excess, enzyme saturation together with an alternating low/high pattern in the site-specific relative catalytic efficiencies, enriches for multistability. That is, within physiologically plausible ranges for parameters, multistability becomes more likely than monostability. We discuss the experimental challenges in pursuing these predictions and in determining the biological role of short-term information storage. \\[0.1em]


\end{abstract}

\clearpage

\section*{Introduction}

Cells are information processing machines which require various forms of information storage, or memory, to carry out their functions. Genetic information, encoded in the structure of DNA, can be stably retained over an evolutionary timescale. Cellular differentiation also requires memory---whether as an adaptive response to the environment by a microbe or as part of the developmental process in a multicellular organism---because information about the differentiated state must be retained stably after the cue which gave rise to it has disappeared. In this case, the memory is implemented not in the structure of a molecule but in the dynamical behaviour of a gene regulatory network \citep{mj61}. Several have now been identified \citep{oud-ea04,gsd04,jceoh05}. The signalling pathways that initiate gene expression may also play a significant role in implementing such memories \citep{bf01,xf03}. These molecular circuits all have similar designs consisting of interlinked positive (or double negative) feedback loops and mathematical analysis shows that they exhibit bistability: they have two stable states, corresponding to two differentiation phenotypes \citep{fer02}. Their information storage capacity, measured in bits, is one. To date, neither experiment nor analysis has shown a multi-bit capacity.

Cells must also possess forms of post-translational short-term memory, for processing external signals. This is most evident in neurons. In hippocampal CA1 cells, transient high-frequency (tetanic) stimulation can enhance a synapse's response to normal stimulation. Such ``long-term potentiation'' (LTP) is thought to underlie neuronal learning and memory \citep{lls03,kan03}. LTP can persist for an hour or more after tetanic stimulation, in a manner independent of protein synthesis (early-phase LTP), while repeated tetanic stimulation results in protein-synthesis dependent synaptic remodelling (late-phase LTP). Early-phase LTP requires a post-translational short-term memory. Crick and Lisman independently suggested the reaction scheme in Figure~\ref{f-rs}A, in which a protein kinase autophosphorylates when activated by single-site phosphorylation \citep{crick84,lisman85}. Lisman's mathematical analysis showed that under phosphatase saturation this positive feedback scheme exhibits bistability. Subsequent work implicated CaMKII as a one-bit molecular memory behind early-phase LTP \citep{lsk02}. This autophosphorylating multimeric kinase is highly concentrated in the post-synaptic density and exemplifies the view that proteins are computational elements which orchestrate cellular information processing \citep{bray95,ks98}. A recent model which presents a synthesis of current data on LTP suggests that a multi-bit capacity may be needed, although an appropriate implementation has not yet been found \citep{lr06}.

Other signal transduction pathways, initiated by hormones, cytokines or growth factors, must also process complex external signals to make appropriate decisions. Engineering theory shows that machines with memory can undertake more complex symbol processing than machines without memory \citep{hmu06}. As capabilities have increased for subjecting cells to complex signals, evidence has grown for post-translational memory mechanisms. Point stimulation of MCF7 cells by beads coated in epidermal growth factor (EGF) results in rapid all-or-none activation of EGF receptors throughout the plasma membrane \citep{vwrb00}. Mathematical and experimental analysis of the double negative feedback loop between EGF receptor activation and tyrosine phosphatase activation by reactive oxygen species shows a bistable mechanism underlying this \citep{rtvrb}.

In this paper we show by mathematical analysis that multisite phosphorylation and dephosphorylation systems, which occur ubiquitously in all eukaryotic signalling pathways, can exhibit many stable states and that the maximal number of steady states increases with the number of sites. The corresponding reaction scheme, which requires no overt positive feedback, is shown in Figure~\ref{f-rs}B. If $n$ is the number of sites, the maximal information storage capacity is at least $\log_2 (n+2)/2 $ bits, when $n$ is even and $\log_2 (n+1)/2$ when $n$ is odd. If the system is initiated with unphosphorylated substrate then, depending on the rate constants, it can reach a different steady state to when the substrate is fully phosphorylated and we give in the Discussion an informal argument to account for this behaviour. Multistability predominates over monostability within physiological ranges, provided substrate is in excess, the kinase and phosphatase are saturated and the site-specific relative catalytic efficiencies follow an alternating low/high pattern. Furthermore, the memory can be switched between stable states by modulating the activity of either kinase or phosphatase. Our results emerge from an analytic solution for the steady state of the system in Figure~\ref{f-rs}B, without the need for any rapid equilibrium or quasi-steady state approximations. 

Multi-bit systems can be built from one-bit systems, as in electronics. However, in the absence of wires and insulation, the number of components required in vivo would scale with the number of bits. Synthetic biologists may hence also be interested in a molecular device with only three components which can store several bits of information \citep{dws07,um06}.

\section*{Results}

\subsubsection*{Preliminary discussion of the model}

We consider a kinase $E$ and a phosphatase $F$ acting distributively and sequentially on a substrate $S$ with $n$ phosphorylation sites. An enzyme acts distributively if it makes at most one modification (addition or removal of phosphate) in each molecular encounter, so that each phospho-form competes for the enzyme. A system is sequential if sites are phosphorylated in a specific order and dephosphorylated in the reverse order. Sequentiality reduces the number of phospho-forms from $2^n$ to $n+1$ and simplifies the analytical treatment developed here. If $S_i$ denotes the phospho-form with $i$ sites phosphorylated in order, then these assumptions lead to the chain of enzymatic reactions in Figure~\ref{f-rs}B. Each enzyme acts through a standard biochemical mechanism, as shown in Figure~\ref{f-rs}B, along with the rate constants appropriate for mass-action kinetics \citep{cb95}. ATP is assumed to be kept constant by some external mechanism, which is not explicitly modelled, and its effect absorbed into the rate constants. 

These assumptions are customary in studies of multisite phosphorylation \citep{hf96,lbs00,sh03,mhk04,gun05,ogmkc} but their relevance to experiment needs to be clarified. Several distributive enzymes have been characterised. Both Mek phosphorylation and MKP3 dephosphorylation of Erk, on two sites, are distributive \citep{fb97,bs97,zz01}, so that the Mek, MKP3, Erk system is an example of a kinase, phosphatase, substrate system that satisfies one of the two assumptions. Sequential kinases have also been characterised. For instance, GSK3, in its primed phosphorylation mode, phosphorylates SXXXS repeat motifs on each serine residue in a strictly C to N order \citep{har01}. FGFR1 has also been shown to autophosphorylate in a strictly sequential manner \citep{flsa06}. Although these observations suggest that cognate phosphatases may act in a similar way, no such phosphatase is currently known. However, unlike distributivity, which is essential for our results, sequentiality is a mathematical convenience. We find that non-sequential systems also exhibit multistability (not shown). We expect this to show the same general properties as for sequential systems, although the maximal number of steady states may be different.

\subsubsection*{The model has an analytic solution for the steady state}

\begin{figure}
\centering 
\includegraphics[viewport=14 14 647 644,width=\textwidth,height=0.5\textheight,keepaspectratio]{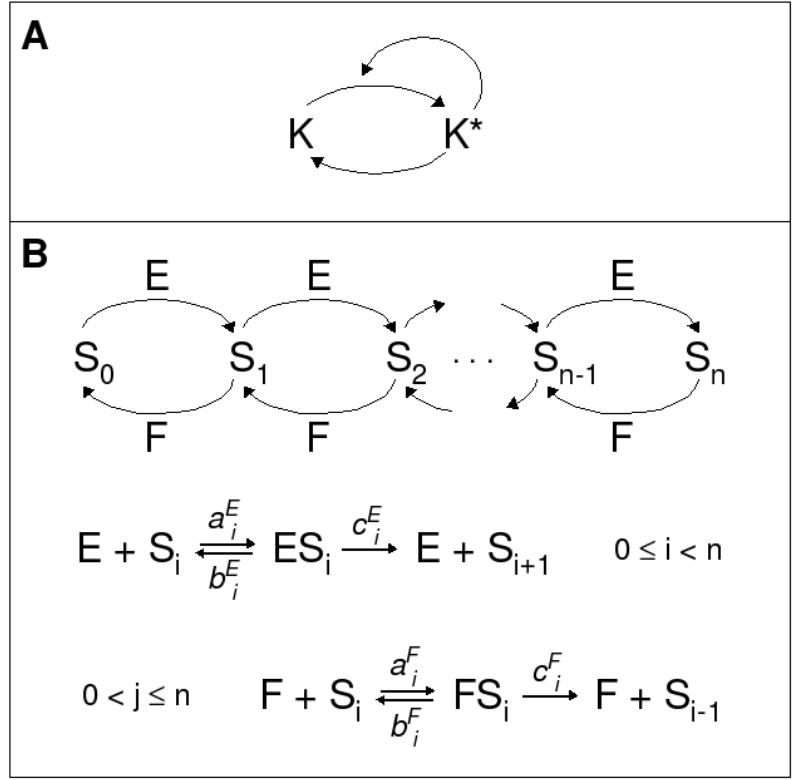}
\caption{Reaction schemes for cellular memory. {\bf A} Lisman scheme \citep{lisman85} in which kinase K autophosphorylates in its active state K*. {\bf B} Scheme considered here, with no explicit positive feedback. Substrate $S$ with $n$ phosphorylation sites is phosphorylated by kinase $E$ and dephosphorylated by phosphatase $F$. Both enzymes act distributively and cooperate to maintain a sequential order. $S_i$ denotes the phospho-form with $i$ sites phosphorylated in sequence. Phospho-forms $S_0, \cdots, S_{n-1}$ have access to $E$ and phospho-forms $S_1, \cdots, S_n$ have access to $F$ through similar reaction schemes, with the reversible formation of enzyme-substrate complexes, $ES_i$ or $FS_j$, respectively, and irreversible formation of product. With mass-action kinetics, each reaction has the indicated rate constant: ({\em a} for ``association''; {\em b} for ``break-up''; {\em c} for ``catalysis''). ATP is assumed held constant and its effect absorbed into the rate constants.\label{f-rs}}
\end{figure}

The reaction scheme in Figure~\ref{f-rs}B gives rise to a dynamical system of $3n+3$ ordinary differential equations which describe the time evolution of $n+1$ phospho-forms, $S_0, \cdots, S_n$; $2n$ enzyme-substrate complexes, $ES_i$ for $0 \leq i < n$ and $FS_j$ for $0 < j \leq n$; and 2 free enzymes, $E$ and $F$. Since the system is closed, the total amounts of substrate, $\st$, and enzymes, $\et, \ft$, are conserved during any time evolution. The system is at steady state if production and consumption of each species is balanced. A steady state is stable if any small perturbation causes a return to the state, as for a ball in a valley; it is unstable if some small perturbation causes the system to run away, as for a ball perched on top of a hill \citep{hs74}.  The system is multistable if there is more than one stable steady state having the same total amounts of enzymes and substrate. The last proviso is important: if the system is initiated with different total amounts of enzymes and substrate then, because the amounts are conserved, it will necessarily find different steady states. This trivial possibility must always be discounted when discussing multistability in systems with conserved quantities.

We showed in previous work \citep{gun05} that this model has an analytic solution at steady state, without the need for rapid equilibrium or Michaelis-Menten or approximations as customarily used \citep{mhk04,ogmkc,sh07}. We briefly recall the argument. Let $[-]$ denote concentration in any steady state. Balancing production and consumption for enzyme-substrate complexes, we find that
\begin{equation}
[XS_i]  =  \frac{[X][S_i]}{K^X_i}\,,
\label{e-es}
\end{equation}
where either $X = E$ and $0 \leq i < n$ or $X = F$ and $0 < i \leq n$. Here, $K^X_i$ denotes the site-specific Michaelis-Menten constant, which, using the notation in Figure~\ref{f-rs}B, is given by
\begin{equation}
K^X_i = \frac{b^X_i + c^X_i}{a^X_i}\,.
\label{e-mmc}
\end{equation}
Now consider the enzymatic chain in Figure~\ref{f-rs}B. If $0 < i < n$, the net flux of substrate into $S_i$ from the left is always equal to the net flux out of $S_{i-1}$ to the right. For all the phospho-forms to be at steady state, it is necessary and sufficient that the net flux into $S_i$ from the left must equal the net flux out of $S_i$ to the right. Since there is never any net flux into $S_0$ from the left or net flux out of $S_n$ to the right (for which sequentiality is essential), it is necessary and sufficient that all the net fluxes are 0. Equivalently, each individual loop in the chain is at steady state. It follows that
\begin{equation}
\frac{[S_{i+1}]}{[S_i]} = \lambda_i \frac{[E]}{[F]} \,,
\label{e-si}
\end{equation}
where $\lambda_i$ is the site-specific relative catalytic efficiency
\begin{equation}
\lambda_i = \left(\frac{c^E_i}{K^E_i}\right)\left(\frac{c^F_{i+1}}{K^F_{i+1}}\right)^{-1} \,.
\label{e-li}
\end{equation}
Applying (\ref{e-li}) repeatedly, we see that
\begin{equation}
[S_{i+1}] = [S_0] \lambda_0\lambda_1 \cdots \lambda_i \left(\frac{[E]}{[F]}\right)^{i+1} \,.
\label{e-si2}
\end{equation}
It follows from (\ref{e-es}) and (\ref{e-si2}) that if the system is at steady state then all $3n+3$ species concentrations are determined by $[S_0], [E]$ and $[F]$. Conversely, if $[S_0], [E]$ and $[F]$ are given arbitrary positive values and the remaining species concentrations are defined by (\ref{e-es}) and (\ref{e-si2}) then it can be readily shown that the system is at steady state. Equations (\ref{e-es}) and (\ref{e-si2}) provide an analytic solution for any steady state of the system in Figure~\ref{f-rs}B.

\subsubsection*{Multiple steady states exist}

As explained above, multistability means the existence of two or more stable steady states having the same total amounts of substrate and enzymes. Equations (\ref{e-es}) and (\ref{e-si2}) enable these total amounts to be calculated in terms of $[S_0], [E]$ and $[F]$. We formalise this in a function $\Phi$, whose properties determine whether or not the system is multistable. To construct $\Phi$, we need to introduce three polynomial functions of $u = [E]/[F]$:
\begin{equation}
\begin{array}{rcl}
\phi_1(u) & = & \displaystyle \sum^n_{i=0} \lambda_0\lambda_1 \cdots \lambda_{i-1} u^i \\[0.1in]
\phi_2(u) & = & \displaystyle \sum^{n-1}_{i=0} \frac{\lambda_0\lambda_1 \cdots \lambda_{i-1}}{K^E_i} u^i \\[0.1in]
\phi_3(u) & = & \displaystyle \sum^n_{i=1} \frac{\lambda_0\lambda_1 \cdots \lambda_{i-1}}{K^F_i} u^i \,.
\end{array}
\label{e-phi}
\end{equation}
These functions have been chosen so that, using (\ref{e-es}) and (\ref{e-si2}), the total amount of substrate is given by (omitting the arguments of the $\phi$ functions for clarity), 
\[
\begin{array}{rcc}
\st & = & [S_0] + \cdots + [S_n] + [ES_0] + \cdots + [ES_{n-1}] + [FS_1] + \cdots + [FS_n] \\[0.5em]
& & = [S_0]\left(\phi_1 + [E]\phi_2 + [F]\phi_3\right) \,,
\end{array}
\]
and, in a similar way, the total amounts of enzymes are
\[
\begin{array}{rcl}
\et & = & [E]\left(1 + [S_0]\phi_2\right) \\
\ft & = & [F]\left(1 + [S_0]\phi_3\right) \,.
\end{array}
\]
Since $\st$ is under the control of the experimenter, while $[S_0]$ is determined by the dynamics of the system, it is preferable to work with $\st$ instead of $[S_0]$, which we can do by using the equation for $\st$. We can then rewrite the equations for $\et$ and $\ft$ in the form of a $2 \times 2$ function, $\Phi$, 
\begin{equation}
\begin{array}{rcl}
\Phi_1([E],[F]) & = & \displaystyle [E]\left(1 + \frac{\st\phi_2}{\phi_1 + [E]\phi_2 + [F]\phi_3}\right) \\
\Phi_2([E],[F]) & = & \displaystyle [F]\left(1 + \frac{\st\phi_3}{\phi_1 + [E]\phi_2 + [F]\phi_3}\right) \,,
\end{array}
\label{e-etft}
\end{equation}
such that $\Phi_1([E],[F]) = \et$ and $\Phi_2([E],[F]) = \ft$. $\st$ has now become part of the definition of $\Phi$. The system is multistable if, and only if, $\Phi$ is many-to-one. In other words, if there are two or more pairs $([E],[F])$ whose $\Phi$ values are the same.

\begin{figure}
\centering 
\includegraphics[viewport=14 14 808 648,width=\textwidth,height=0.5\textheight,keepaspectratio]{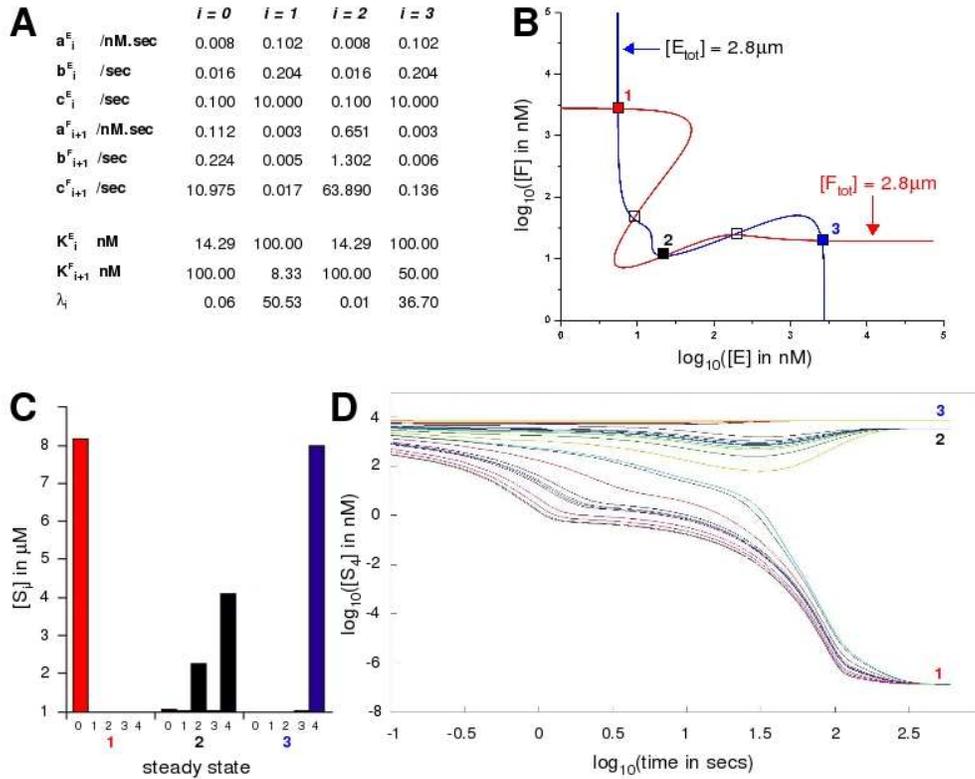}
\caption{Multiple steady states for a system with four phosphorylation sites. {\bf A} Rate constants, rounded to three decimal places. $K^E_i$, $K^F_i$ and $\lambda_i$ are needed to determine the steady states; the other rate constants are needed to determine stability. {\bf B} Plots of $\Phi_1([E],[F]) = 2.8\,\uM$ and $\Phi_2([E],[F]) = 2.8\,\uM$ with $\st = 10\,\uM$, showing five steady states at the intersections. Filled squares are stable and labelled 1 (red), 2 (black) and 3 (blue); open squares unstable. Log scales on both axes. {\bf C} Bar chart of $[S_0], \cdots, [S_4]$ in $\uM$, labelled by subscript on the horizontal axis, for each of the three stable states, as previously labelled. {\bf D} Time courses of $S_4$ reaching its three different stable values from initial conditions $[S_0] = \alpha\st$, $[S_4] = (1-\alpha)\st$ and $[X] = 0$ for all other species $X$, with $\alpha$ chosen randomly in $[0,1]$, obtained by model simulation. Log scales on both axes. \label{f-ex}}
\end{figure}

Suppose that rate constants are determined and the total amounts of substrate and enzymes are chosen, $\et = A, \ft = B, \st = C$. To determine whether or not the system is multistable, it suffices to solve the pair of equations
\begin{equation}
\Phi_1([E],[F]) = A\,, \hspace{1em} \Phi_2([E],[F]) = B 
\label{e-neq}
\end{equation}
simultaneously for $[E]$ and $[F]$. This can be done numerically as described in Materials and Methods. The solutions give all the steady states of the system for which $\et = A, \ft = B$ and $\st = C$. The curves defined by (\ref{e-neq}) may also be plotted in the $([E],[F])$ plane where their intersections show the steady states. Figure~\ref{f-ex}B gives an example with four sites. The corresponding rate constants in Figure~\ref{f-ex}A appear physiologically plausible, given our current limited understanding of site-specific rate constants. The curves for $\et = 2.8\,\uM$, $\ft = 2.8\,\uM$ and $\st = 10\,\uM$ have five intersections, giving five steady states. A separate analysis shows that three are stable and two unstable, as indicated. These stable states have widely different mixtures of the phospho-forms, as shown in Figure~\ref{f-ex}C. We simulated the corresponding dynamical system and found that unphosphorylated substrate reached the steady state with low $[E]$ and high $[F]$, fully phosphorylated substrate reached the state with high $[E]$ and low $[F]$ (the ``outer'' states) and a suitable mixture of phospho-forms reached the inner state, as shown in Figure~\ref{f-ex}D. These behaviours were characteristic of the multistable systems we simulated and provide a method for detecting multistability experimentally. We give an informal explanation for the outer states in the Discussion.

Bistability was first shown for $n = 2$ in \citep{mhk04}. It was later claimed that no more than two stable states occur when $n > 2$ \citep{ogmkc}. This is incorrect, as we have just shown.

\subsubsection*{A simplified solution exists when substrate is in excess}

$\Phi$ gives an exact solution in two dimensions for the steady states of a $3n+3$-dimensional dynamical system. However, numerical solution of (\ref{e-neq}) is computationally expensive. It can take up to thirty seconds to find all the steady states for a system with four sites, making it difficult to explore the conditions under which multistability arises. We found by exploration that multistability occurs when substrate is in excess so we considered what happens when either enzymes or substrate are in excess. If enzymes are in excess, enzyme-substrate complexes are negligible in comparison to $\et$ and $\ft$. Hence, $\et \approx [E]$ and $\ft \approx [F]$, $\Phi$ is one-to-one and the system is monostable \citep{gun05}. If substrate is in excess, then the total amounts of enzyme-substrate complexes may be considered negligible in comparison to $\st$. Hence, we may write, approximately,
\[ \st = [S_0] + \cdots + [S_n] = [S_0]\phi_1(u) \,,\]
where $u = [E]/[F]$. We can then rewrite (\ref{e-etft}) to get
\begin{equation}
\begin{array}{rcl}
\et & = & \displaystyle [E]\left(1 + \st\frac{\phi_2(u)}{\phi_1(u)}\right) \\[1em]
\ft & = & \displaystyle [F]\left(1 + \st\frac{\phi_3(u)}{\phi_1(u)}\right) \,.
\end{array}
\label{e-xs}
\end{equation}
For given $\et, \ft$ and $\st$, the $([E],[F])$ pairs which are solutions of (\ref{e-xs}) are the steady states of the system, to within the approximation. Dividing the first equation by the second, and setting $\et/\ft = w$, we see that
\[ w(\phi_1(u) + \st\phi_3(u)) = u(\phi_1(u) + \st\phi_2(u)) \,,\]
and so, rearranging this, 
\begin{equation}
P(u) = (u - w)\phi_1(u) + \st(u\phi_2(u) - w\phi_3(u)) = 0 \,.
\label{e-df2}
\end{equation}
Since $\phi_1(u)$, $\phi_2(u)$ and $\phi_3(u)$ are all polynomial functions of $u$, $P(u)$ is a polynomial function of $u$, whose degree is $n+1$. 

For each $([E],[F])$ pair which is a solution to (\ref{e-xs}), $u = [E]/[F]$ is a positive solution of $P(u) = 0$. It can be checked that the converse is also true. Hence, solutions of the approximate system (\ref{e-xs}) correspond exactly to positive roots of $P(u)$. Suppose that
\begin{equation}
P(u) = a_{n+1}u^{n+1} + a_n u^n + \cdots + a_1 u + a_0 \,.
\label{e-poly}
\end{equation}
The coefficients $a_i$ may be calculated from (\ref{e-df2}):
\begin{equation}
\begin{array}{rcll}
a_{n+1} & = & \lambda_0\cdots\lambda_{n-1},\;\;\;a_0 = -w \;\;\;\mbox{and, for $0 \leq i < n$,}\\[0.4em]
a_{i+1} & = & \displaystyle \lambda_0\cdots\lambda_{i-1}\left[(1 - \lambda_i w) + \st\left(\frac{1}{K^E_i} - \frac{\lambda_i w}{K^F_{i+1}}\right)\right] \,.
\end{array}
\label{e-coef}
\end{equation}

Polynomial root finding is computationally fast and we will use this to search for steady states. We conducted tests and chose $\st/\et \geq 5$ as the limit for these searches. In this range, the average normalised difference between the solution values reported by $\Phi$ and by $P(u)$ is at most $0.23$, as shown in Figure~\ref{f-apx} and explained further in Materials and Methods. The frequency of potential miscounting of steady states by $P(u)$ is $0.2\%$ (7/3385). We considered these rates acceptable for the random searches below.

\subsubsection*{The information storage capacity is at least $\log_2\lfloor (n+2)/2 \rfloor$ bits}

A polynomial of degree $n+1$ has at most $n+1$ roots \citep{hs74}. However, only positive roots are relevant for us. Descartes' Rule of Signs \citep{ajs98} states that the number of sign changes in the coefficients of $P(u)$ exceeds the number of its positive roots by a non-negative even integer. We know from (\ref{e-coef}) that $a_{n+1} > 0$ and $a_0 < 0$. Hence, if $n$ is odd, there can be at most $n$ sign changes, while if $n$ is even, there can be at most $n+1$ sign changes:
\[ n = 3: \; \overbrace{+ - + - -}^{\mbox{3 sign changes}} \hspace{3em} n = 4: \; \overbrace{+ - + - + -}^{\mbox{5 sign changes}} \,.\]
Accordingly, if $n$ is odd, the maximum number of steady states is $n$, while if $n$ is even, the maximum is $n+1$. These bounds are attained because we can show that any polynomial like (\ref{e-poly}) for which $a_{n+1} > 0$ and $a_0 < 0$ can be obtained by arbitrary choice of $\st > 0$ and appropriate choice of $K^E_i$, $K^F_i$, $\lambda_i$, $w = \et/\ft$ all positive, in (\ref{e-coef}). In particular, this can be done in such a way as to ensure that the approximation to the exact system (\ref{e-neq}) is as close as required. The details are given in Materials and Methods. 

Suppose then that $n$ is odd and $\alpha_1, \cdots \alpha_n$ are any $n$ distinct positive numbers. The polynomial $(u - \alpha_1)(u - \alpha_2) \cdots (u - \alpha_n)(u + 1)$ has degree $n+1$ and satisfies $a_{n+1} > 0$ and $a_0 < 0$. Similarly, if $n $ is even and $\alpha_1, \cdots, \alpha_{n+1}$ are any $n+1$ distinct positive numbers, the polynomial $(u - \alpha_1) \cdots (u - \alpha_{n+1})$ has degree $n+1$ and also satisfies these conditions. Hence, not only can we find rate constants for which the above upper bounds are attained, we can also ensure that the values of $u = [E]/[F]$ at the steady states are any arbitrary pre-assigned distinct positive numbers. It is possible that, outside the range of approximation, the system has more steady states than positive roots of $P(u)$. \citep{wsX704}, following on from our results, have used singular perturbation theory to show that there are not more than $2n$ steady states. However, we conjecture that the bounds established in this paper always hold.

On the basis of separate tests for stability, as discussed in Materials and Methods, we concluded that the number of stable steady states is $\lfloor (n+2)/2 \rfloor$. Since the information storage capacity of a system with $k$ stable states is $\log_2 k$ bits, the maximal information storage capacity is at least $\log_2\lfloor (n+2)/2 \rfloor$ bits. Multisite phosphorylation and dephosphorylation systems are capable of multi-bit information storage whose maximum capacity increases with the number of sites.

\subsubsection*{An alternating low/high pattern of $\lambda_i$ enriches for multistability}

Under what conditions on rate constants and amounts does multistability occur and are these physiologically plausible? As just seen, the first question is related to when a polynomial has many positive roots. We found this to be mathematically intractable, as explained in Materials and Methods. Indeed, only probabilistic answers have been found to this general class of questions. For instance, if the coefficients of (\ref{e-poly}) are chosen randomly from the standard normal distribution, the average number of real roots (ie: without restriction on the sign) is given by the Kac integral formula, which is approximated by $2\log (n+1)/\pi$ \citep{ek95}. Proportionately, very few of the roots of a random polynomial are real; for a random polynomial of degree 100, the average number of real roots is only 3.56 \citep{brs86}. This suggests that high multistability, while mathematically possible, is exceedingly rare. However, it still leaves open the possibility that some bias in the coefficients can enrich for it.

According to the Rule of Signs, the number of positive roots of $P(u)$ can only reach its maximum value of $n+1$ when the number of sign changes in the coefficients is as high as possible. The sign of $a_{i+1}$, as given by (\ref{e-coef}), is the net result of two additive terms, each of which may be positive or negative. We will re-interpret these terms in the Discussion but in the special case when $K^E_i = K^F_{i+1}$, the sign of $a_i$ is determined solely by $(1 - \lambda_i w)$. Hence, for maximum sign changes, the $\lambda_i$ should satisfy an alternating low/high pattern (assuming $n$ even):
\begin{equation}
\lambda_0 < \frac{1}{w}, \hspace{1em} \lambda_1 > \frac{1}{w}, \hspace{1em} \cdots, \hspace{1em} \lambda_{n-1} > \frac{1}{w} \,.
\label{e-alt}
\end{equation}
We found this pattern in many examples with high multistability, like the system in Figure~\ref{f-ex}. It is not equivalent to the alternating sign condition but has the merit of only involving one of the parameters.

We find that (\ref{e-alt}) enriches for multistability. We take a probabilistic approach to demonstrating this, in the light of the mathematical results mentioned above. For each even $n$ from 2 to 12 we generated 100,000 systems as follows. We chose $\log_{10}(K^X_i\,\mbox{ in $\nM$})$ randomly from the uniform distribution on $[-1,2]$ and $\log_{10}\lambda_i$ randomly from the uniform distribution on $[-2, 2]$. We set $\st = 1000\,\nM$, forcing the enzymes into saturation, and $\et = \ft = 200\,\nM$, ensuring that substrate was in excess. We found the distribution of steady states in Figure~\ref{f-ms}A where monostability remains more likely than multistability up to $n = 12$ and five steady states do not appear until $n = 6$. We then repeated the calculation with $\log_{10}\lambda_i$ uniform on $[-2,0]$ for $n$ even and on $[0,2]$ for $n$ odd, following the alternating low/high pattern described by (\ref{e-alt}), with $w = 1$. The distribution shifted to that in Figure~\ref{f-ms}B in which multistability is now more likely than monostability as soon as $n > 2$, the frequency of five steady states is increased and becomes non-zero for $n = 4$. Saturation plays an important role here. We took $\st = 10\,\uM$ and $\et = \ft = 2\,\uM$ and found that monostability is now overwhelmingly more likely and that the bias in $\lambda_i$ has much less effect (data not shown). Hence, within physiologically plausible ranges, substrate excess, saturated enzymes and an alternating low/high pattern in the relative catalytic efficiencies enriches for multistability.

\subsubsection*{Modulating enzyme activity leads to hysteresis}

If a multisite protein phosphorylation system acts as a memory device, it is unlikely to be regulated in vivo by altering its initial condition. It is more plausible that the activity of one of the enzymes will be modulated. We simulated the dynamical system in Figure~\ref{f-ex}, taking it through a cycle in $\et$ by changing free kinase a small amount and letting the system relax back to a steady state after each perturbation. We found hysteresis, as shown in Figure~\ref{f-hyst}A. As $\et$ is increased the system reaches a bifurcation point \citep{hs74} where it jumps abruptly to a higher branch; when $\et$ is then reduced, the system remains on the higher branch beyond the bifurcation point, until jumping down to a lower branch at a lower value of $\et$. $\et$ can therefore be cycled and the system switched between the outer states in Figure~\ref{f-ex}A. Modulation of the enzymes can rewrite the memory. This provides another method for detecting multistability experimentally, which is more feasible than altering rate constants to show hysteresis.

Surprisingly, systems with fewer steady states can show more complex hysteresis. When $\st$ is reduced to $5\,\uM$ the system in Figure~\ref{f-ex}B becomes bistable with only three steady states. However, a similar cycle in $\et$ produces the double hysteresis in Figure~\ref{f-hyst}B, showing that the system finds three stable states even though there is only a narrow window for $\et$ in which three stable states exist simultaneously. The potential for it, however, affects the complexity of hysteresis. We found a similar effect in the approach to steady state (data not shown). When a system is close in parameter space to regions of higher multistability, these nearby stable states can exert a complex influence on the dynamics. When there is merely the potential for higher multistability, as, for instance, when $n$ is large, the dynamic and hysteretic behaviour of a system may reflect that complexity, even though the number of steady states in the actual system is low.

\section*{Discussion}

\subsubsection*{Summary}

We have shown that a system with three molecular components, a kinase, a phosphatase and a substrate with $n$ phosphorylation sites, can exhibit multiple stable steady states and thereby function as a multi-bit post-translational cellular memory. The maximum information capacity increases with increasing numbers of sites and is at least $\log_2\lfloor (n+2)/2 \rfloor$ bits. The conditions on rate constants for multistability to exist are mathematically intractable but, when substrate is in excess, enzyme saturation together with an alternating low/high pattern in the site-specific relative catalytic efficiencies enriches for multistability. That is, when rate constants are taken within physiological ranges and randomly sampled as specified above, multistability becomes more likely than monostability as soon as $n > 2$. The different states of the memory can be selected by modulating the activity of one of the enzymes. Even if a system has low multistability relative to the maximum, its dynamic and hysteretic behaviour can show the influence of nearby regions of parameter space with higher multistability. Our results suggest two methods for detecting multistability: different mixtures of phospho-forms can pick out different steady states---in particular, unphosphorylated substrate and fully phosphorylated substrate can pick out the outer steady states---while enzyme cycling can show hysteresis.

While these results have been framed for protein phosphorylation and dephosphorylation systems, they are potentially applicable to any reversible modification, such as protein ubiquitination or histone methylation \citep{rrk06,kz07}, that follows a similar scheme to Figure~\ref{f-rs}B. However, much less is known about multisite effects in such systems.

\subsubsection*{Multistability through kinetic trapping}

We can provide some intuition at to why the three conditions of substrate excess, enzyme saturation and low/high pattern of $\lambda_i$ give rise to two outer steady states. By ``outer'', we mean those steady states which have minimum or maximum $[E]/[F]$ value; all other steady states are ``inner''. Unlike the steady-state analysis presented above, the argument given here follows the dynamics of the system from a given initial condition. In contrast to the steady state, the dynamics does not have an analytic solution, hence our argument is an informal one.

Suppose that a multisite system has substrate in excess over enzymes and that the total amount of substrate saturates both kinase and phosphatase at each site. These are two of the three conditions. Let us start the system in state $S_0$ with all the substrate unphosphorylated. Since $E$ is saturated by $S_0$, the rate of production of $S_1$ will immediately reach a near maximal value, which will remain nearly constant as long as $S_0$ continues to saturate $E$. As $S_1$ is produced, it will become available to both $E$, to produce $S_2$, and $F$, to produce $S_0$. However, the former reaction will be negligible because $S_0$, being in excess, will have sequestered free enzyme away from $S_1$. The latter reaction, however, will proceed, as $F$ is unoccupied. What happens next depends on the relative behaviour of $E$ and $F$ acting in the loop between $S_0$ and $S_1$. Let us assume that both enzymes work approximately according to the Michaelis-Menten rate law and recall \citep{cb95} that these take the form
\begin{equation}
\frac{c^E_0 \et [S_0]}{K^E_0 + [S_0]} \hspace{1em}\mbox{and}\hspace{1em} \frac{c^F_1 \ft [S_1]}{K^F_1 + [S_1]}\,.
\label{e-mm}
\end{equation}
Finally, consider a third condition: suppose that the rate curve for $F$ lies entirely above that for $E$, as shown in Figure~\ref{f-intuit}. We will interpret this in terms of the low/high pattern below. In this arrangement of the curves, the rate of production of $S_0$ from $S_1$ by $F$ can rapidly rise until it meets the nearly maximal rate of production of $S_1$ from $S_0$ by $E$, at which point the $S_0$ to $S_1$ loop will be in steady state. Although there might be a leak from $S_1$ to $S_2$, this will be small, as long as $S_0$ is in excess, and will be immediately balanced by back flow from $S_2$ to $S_1$, since $F$ is not sequestered. Hence, it seems plausible that the system will come to steady state with a substantial amount of $S_0$, a much smaller amount of $S_1$ and very little else. The phospho-form distribution becomes trapped at one end of the chain. Note that no other arrangement of the curves will give such trapping. If the same conditions are applied to the other end but reversed with respect to $E$ and $F$, then fully phosphorylated substrate will become trapped predominantly as $S_n$ and the system will have at least two steady states. The two outer steady states in Figure~\ref{f-ex}C show exactly the distribution of phospho-forms suggested here.

The third condition requires that, first, the initial slope of the $F$ curve at zero substrate exceeds that for the $E$ curve and, second, that the maximal (asymptotic) value of the $F$ curve also exceeds the maximal value for the $E$ curve. From (\ref{e-mm}) these correspond to
\[ \frac{c^E_0 \et}{K^E_0} < \frac{c^F_1 \ft}{K^F_1} \hspace{1em}\mbox{and}\hspace{1em} c^E_0 \et < c^F_1 \ft \,\]
respectively, which may be rewritten as
\[ 1 - \lambda_0 w > 0 \hspace{1em}\mbox{and}\hspace{1em} \frac{1}{K^E_0} - \frac{\lambda_0 w}{K^F_1} > 0 \,,\]
where, as previously, $w = \et/\ft$. We see from (\ref{e-coef}) that this forces the coefficient $a_1$ of $P(u)$ to be positive. For the other end of the chain, we get the opposite effect, with $a_n < 0$. In other words, we recover the outer terms of the alternating sign condition that we found above as a necessary condition for multistability, from which (\ref{e-alt}) emerges as a special case. We see, furthermore, that the two additive terms in the expression for $a_{i+1}$ in (\ref{e-coef}) can be interpreted in terms of the arrangement of the Michaelis-Menten curves for $E$ and $F$ acting between $S_i$ and $S_{i+1}$. The particular arrangement in Figure~\ref{f-intuit} fixes the sign of $a_{i+1}$. 

This informal argument cannot be easily extended to the inner states. If substrate is prepared in an intermediate state of phosphorylation, $S_i$, where $0 < i < n$, then both $E$ and $F$ become sequestered and saturated immediately. Substrate will accumulate as $S_{i-1}$ and $S_{i+1}$ until one or or both of $E$ and $F$ become accessible to other phospho-forms. Which of these happens will depend on other rate constants like $a^X_i$ and $b^X_i$, which determine the dynamics, and not just on the ones which determine the steady state like $K^X_i$ and $\lambda_i$ (or, alternatively, $K^X_i$ and $c^X_i$). Hence, there will be many routes through which a steady state is attained, making any further informal analysis challenging. The intractability of the mathematical conditions for multistability presumably reflects this complexity. Nevertheless, the inner steady state in Figure~\ref{f-ex}C has an unusual distribution, with substrate concentrated predominantly in even numbered phospho-forms, suggesting that a similar type of ``kinetic trapping'' continues to determine the phospho-form distribution.

\subsubsection*{Emergent complexity in phosphorylation, dephosphorylation systems}

Phosphorylation and dephosphorylation are ubiquitous and fundamental regulatory processes, which occur in all organisms. It used to be thought that prokaryotes and eukaryotes used fundamentally different phosphorylation chemistries but a closer look has revealed a more nuanced picture. Bacteria predominantly, but not exclusively, use the two-component histidine, aspartate phospho-transfer process, while eukaryotes predominantly, but again not exclusively, rely on serine, threonine and tyrosine phosphorylation \citep{ken03,srg00,mogsjp}. A more significant distinction between the two kingdoms may be the extent of multisite modification. Two-component systems typically have a single phosphorylation site on the sensor and the response-regulator. A recent analysis of serine, threonine, tyrosine phosphorylation in Bacillus subtilis reveals a few proteins with five to eight phosphorylation sites \citep{mmogkjm} and similar maximum numbers are emerging from further bacterial studies (Boris Macek, personal communication, 2007). Eukaryotic proteins, however, can be far more heavily phosphorylated: p53, for instance, has at least 16 sites which are known to have regulatory function \citep{htes02}.

Many suggestions have been made to account for multisite modification: signal integration, complex logic, attachment points to assemble signalling complexes, structural change through electrostatic effects, allovalency, etc \citep{coh00,htes02,kpt03}. While these may all be relevant, it is still puzzling why quite so many sites are needed. A single substrate molecule with $n$ sites may, in principle, occupy $2^n$ states (over 4000 for p53) and a population of such molecules will exhibit a distribution of these phospho-forms. It is not clear how such complexity can be effectively regulated \citep{gun05,sh07}. Moreover, the system of kinases, phosphatases and substrate is maintained far from equilibrium in vivo by a steady supply of ATP. This is a recipe for complex emergent behaviour, as our mathematical results suggest. The in-vitro reconstitution of a cyanobacterial circadian oscillator \citep{niinmiok}, which manifests itself as an oscillation in multisite phosphorylation, may be an instance of such emergent complexity but it has otherwise proved difficult to study experimentally.

\subsubsection*{Experimental detection of multistability}

We argued in the Introduction that signal transduction systems may require post-translational information storage in order to interpret complex external signals. If so, neither the storage mechanism nor its functional significance may be experimentally detectable in vivo without the ability to control and manipulate the signals. This is clear from studies of LTP in neurons: without tetanic stimulation, or some other complex signal to induce LTP, there would be no memory process to observe. T-cell activation is another context where information processing tasks have begun to be characterised on the basis of their response to complex signals. The T cell receptor is capable of being both highly discriminating among antigens and highly sensitive to small amounts of antigen and can accomplish both tasks quickly, a feat which requires an intricate mixture of kinetic proof-reading and feedback \citep{ag05}. It would not be a surprise to find short-term memory requirements in this kind of immunological synapse as well. While it is technically more difficult to create and control signals from growth factors, cytokines or hormones, the use of microfluidic devices is bringing about a substantial improvement in such experimental capabilities \citep{mq07}. 

The major technical obstacle in vivo, however, is the need for single-cell resolution. If multistability exists, different cells in a population may be in different steady states and a population average could smear out the very signals that are being sought \citep{fm98,lrsglea}. While single-cell sensors of kinase activity have been developed \citep{rsi05}, it remains challenging to determine phosphorylation state in individual cells.

In-vitro studies are more feasible but, outside of extract systems \citep{kf07}, kinases and phosphatases have usually been studied separately (and the former more so than the latter). Steady states, in which kinase and phosphatase are opposed, have not been analysed, although there is no impediment to doing so. Care may be needed to ensure that the ATP is kept in sufficient excess and that ADP build-up does not compromise the reaction scheme in Figure~\ref{f-rs}B. Continuous-flow ATP regenerating systems, as used for in-vitro translation, may help \citep{sbroa,kc96}. The main difficulty lies in distinguishing and quantifying all $2^n$ phospho-forms of a substrate with $n$ sites. Antibodies can be highly selective but we have found that, even for a well-studied substrate like Erk with only two sites, commercial antibodies against intermediate phospho-forms show too much cross-reactivity for accurate quantitation. Phospho-peptide mapping by chromatographic or electrophoretic separation has been successful for low numbers of sites \citep{fb97,zz01} but mass spectrometry is now the proteomic method of choice and shows much promise for phospho-protein analysis \citep{mogsjp,sjrmk}. In collaboration with Hanno Steen, we are developing methods for resolving and quantifying all $2^n$ phospho-forms using a combination of iso-electric focussing, HPLC and mass spectrometry. If multistability in multisite phosphorylation can be detected in vitro, it seems likely that nature will have exploited it in vivo.

\section*{Materials and methods}

\small

\subsubsection*{Numerical solution of $\Phi$}

If $K^E_i$, $K^F_i$ and $\lambda_i$ are specified and $\et = A$, $\ft = B$ and $\st = C$ are chosen, then (\ref{e-neq}) is solved numerically in Matlab (The MathWorks, Natick, MA, USA) as follows. We first calculate $\Phi$ on a grid in the $([E], [F])$ plane and use {\tt contourc} on the output to determine the sets of points satisfying $\Phi_1([E], [F]) = A$ (the $\et$ curve) and $\Phi_2([E], [F]) = B$ (the $\ft$ curve). {\tt Contourc} interpolates to find these ``isolines''. They provide the visual plots in which the steady states appear at the intersections of the curves, as in Figure~2A of the paper. For automated searches we use a $120 \times 120$ grid, where $\log_{10}$ of each coordinate is equally spaced in $[-6,6]$. For manual inspection at finer resolution we use a $1200 \times 1200$ grid. We then calculate the steady states via {\tt fsolve}, which uses an iterative nonlinear search starting from a specified initial condition. We separately calculate the derivatives of $\Phi$ (the Jacobian) and provide that to {\tt fsolve} to speed up the search. An appropriate choice of initial conditions is essential for both speed and accuracy. We found that points lying on either the $\et$ curve or the $\ft$ curve provided good initial conditions, while other points sometimes caused {\tt fsolve} to diverge or return an error. We used the $\et$ curve for the set of initial conditions. We first chose three points on the $\et$ curve, one each at either extreme of $[E]/[F]$ value and the third in the middle. If, for each of these initial conditions, {\tt fsolve} returns a solution and the solutions agree to within a specified tolerance (usually $10^{-4}$) in each coordinate, we return that solution as the unique steady state of the system. If any of these conditions fails, we take every other point lying on the $\et$ curve and run {\tt fsolve} on all of them. We count the resulting solutions as distinct if they differ by more than the tolerance in any coordinate. The distinct solutions are returned as the steady states. This protocol was fine-tuned from numerical experiments to provide a reasonable balance between speed and accuracy, using the visual plot and the numerical calculation to cross-check each other. It can still take up to 30 seconds to find all the steady states for a system with four sites.

\subsubsection*{Stability of steady states}

A dynamical system is defined by a system of ordinary differential equations, $dx/dt = f(x)$, where $x \in \R^m$ and $f: \R^m \ra \R^m$. The Jacobian matrix, $J$, is given by $J_{ij} = \partial f_i/\partial x_j$. According to standard theory, the stability of a steady state is determined by the eigenvalues of the Jacobian evaluated at the state \citep{hs74}. If all the eigenvalues have negative real part, the state is stable; if not, it is unstable. We computed the Jacobian symbolically in terms of the rate constants $a^X_i, b^X_i, c^X_i$ and the steady-state species concentrations $[Y]$. For a given steady state defined by $\st, [E], [F]$, we computed all the steady-state species concentrations using (\ref{e-es}) and (\ref{e-si2}), as described above, and substituted these values into the symbolic Jacobian along with the rate constants. We then calculated the eigenvalues using Matlab's {\tt eig} function. Because the total amounts of substrate and enzymes are conserved we ignored the three resulting zero eigenvalues in determining the stability of a steady state. We found that the other eigenvalues depended on all the rate constants and not just on $K^X_i$ and $\lambda_i$, which determine the steady state.

In tests of stability we found that if the steady states are ordered by increasing $[E]/[F]$, unstable states typically occur between stable ones, so that typically there are $(n+2)/2$ stable states if $n$ is even and $(n+1)/2$ stable states if $n$ is odd. Both cases are covered by $\lfloor (n+2)/2 \rfloor$, where $\lfloor x \rfloor$ denotes the greatest integer not greater than $x$.

\subsubsection*{Approximation of $\Phi$ by $P(u)$}

\begin{figure}
\centering 
\includegraphics[viewport=14 14 797 326,width=\textwidth,height=0.5\textheight,keepaspectratio]{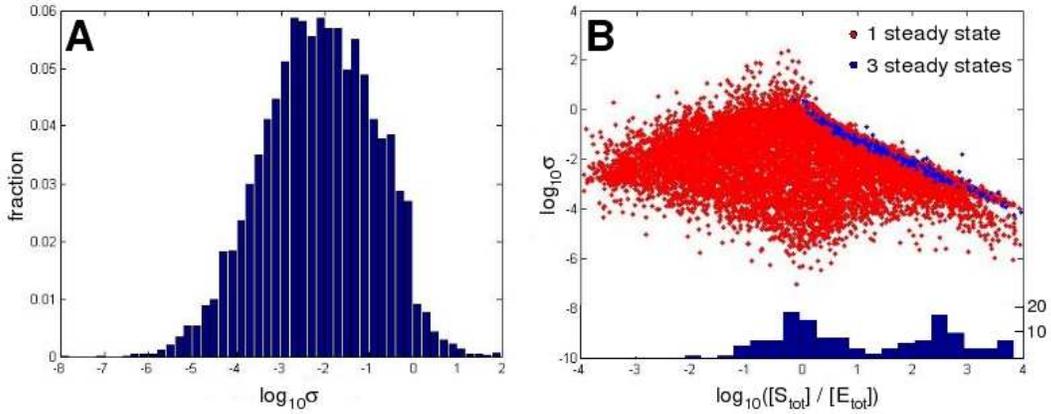}
\caption{Approximation of $\Phi$ by $P(u)$. 10000 random systems were generated, as described in the text, and solved using $\Phi$ and $P(u)=0$. For those which gave the same number of steady states, the discrepancy between the solutions was measured using $\sigma$, as described in the text. {\bf A} Histogram of $\log_{10}\sigma$ values. {\bf B} The top shows a scatter plot of $\log_{10}\sigma$ on the left vertical axis against $\log_{10}\st/\et$. The bottom shows the number of systems which gave different numbers of steady states for $\Phi$ and $P(u)$, using the lower part of the right vertical axis, binned against $\log_{10}\st/\et$.\label{f-apx}}
\end{figure}

\begin{figure}
\centering 
\includegraphics[viewport=14 14 738 297,width=\textwidth,height=0.5\textheight,keepaspectratio]{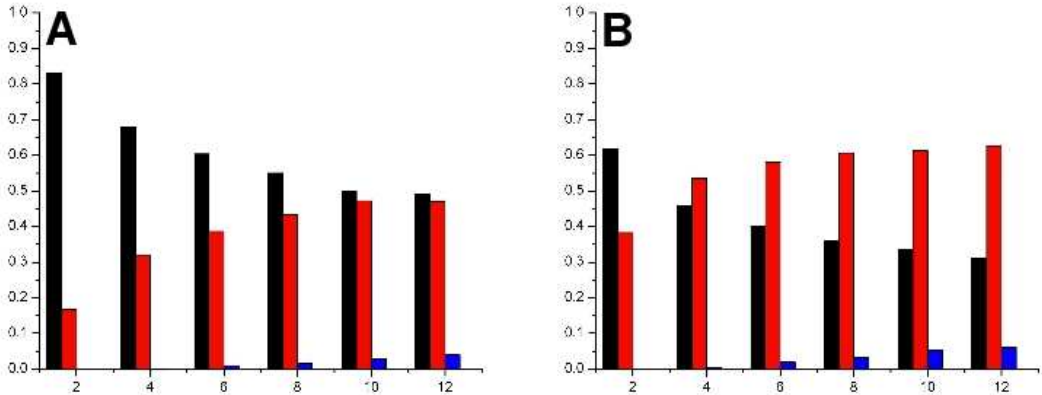}
\caption{Frequency distributions of steady states for randomly chosen systems with $n = 2$ to 12 sites, as described in the text. {\bf A} $\lambda_i$ is chosen uniformly from site to site. {\bf B} $\lambda_i$ is biased to be low for even $i$ and high for odd $i$. Vertical scales show frequency of occurrence of 1 (black), 3 (red) and 5 (blue) steady states, for 100,000 systems for each $n$. \label{f-ms}}
\end{figure}

\begin{figure}
\centering 
\includegraphics[viewport=14 14 481 543,width=\textwidth,height=0.5\textheight,keepaspectratio]{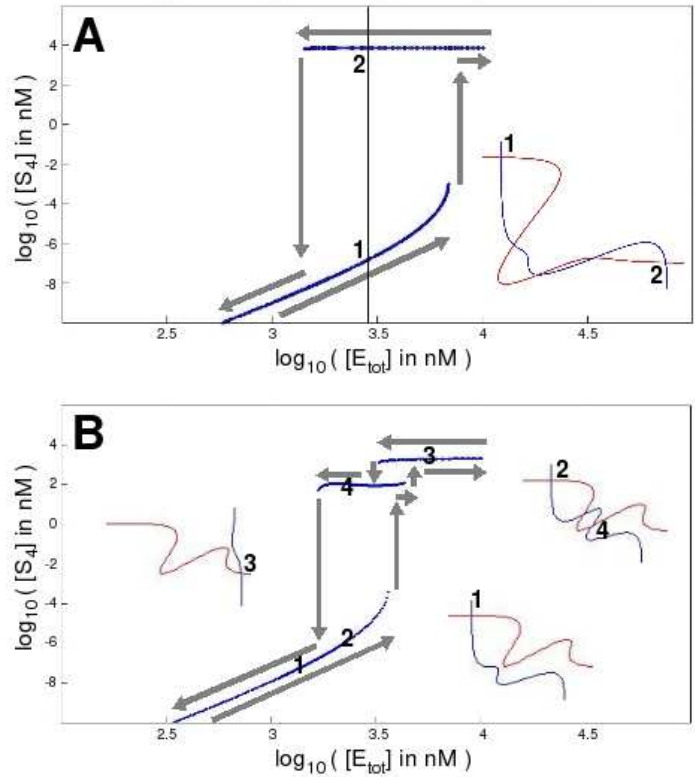}
\caption{Hysteresis for the system in Figure~\ref{f-ex}A. $\et$ is taking in a cycle shown by the grey arrows, as described in the text. Note log scales on both axes. {\bf A} The system with $\et = \ft = 2.8\,\uM$ and $\st = 10\,\uM$, as in Figure~\ref{f-ex}B, has 3 stable states. The vertical line shows its position and the numbers 1-2 mark its positions on the two branches of the cycle and label the corresponding steady states on the inserted $([E],[F])$ plot from Figure~\ref{f-ex}A. {\bf B} The same system with $\st = 5\,\uM$ has only two stable states but occupies three during the cycle. The numbers 1-4 mark positions along the cycle---for 1, $\et = 1.41\,\uM$; for 2 and 4, $\et = 2.04\,\uM$; for 3, $\et = 5.37\,\uM$---and also the corresponding steady states on the $([E],[F])$ curve inserts. Changing $\et$ alters the $\et$ curve but keeps the $\ft$ curve fixed. \label{f-hyst}}
\end{figure}

\begin{figure}
\centering 
\includegraphics[viewport=14 14 474 324,width=\textwidth,height=0.4\textheight,keepaspectratio]{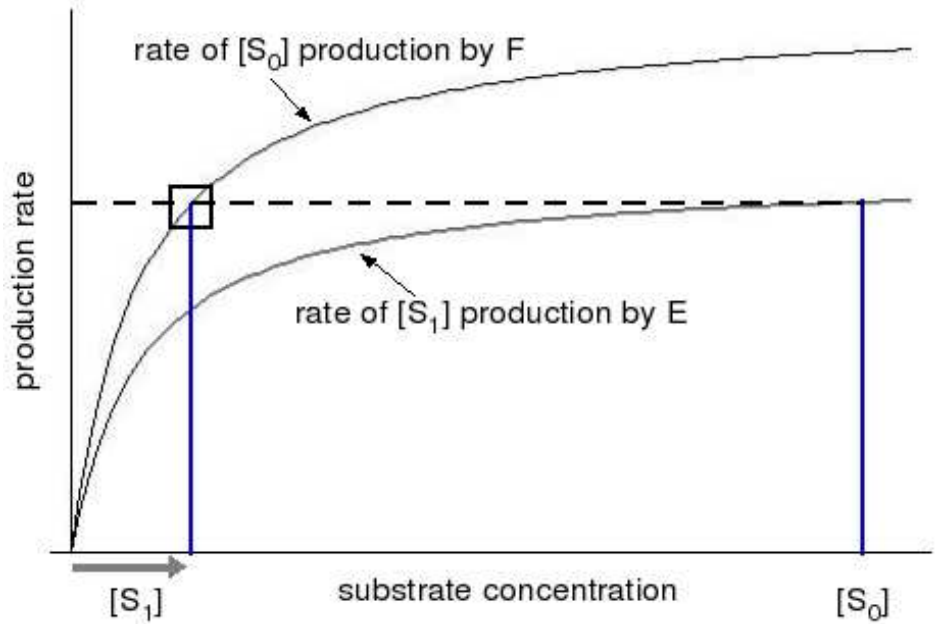}
\caption{Intuitive explanation for the outer steady states. Michaelis-Menten rate curves are shown for $E$ producing $S_1$ from $S_0$ (lower curve) and $F$ producing $S_0$ from $S_1$ (upper curve). The system is started with substrate unphosphorylated in state $S_0$, so that $[S_0]$ is high, as shown, and $S_1$ is produced at a nearly maximal and constant rate, indicated by the dotted line. $F$ is unoccupied and $[S_1]$ rapidly increases (grey arrow) until the rates balance, indicated by the open square, giving rise to a steady state. This arrangement of the curves leads to $a_1 > 0$ in $P(u)$, as explained in the text. \label{f-intuit}}
\end{figure}

To assess quantitatively how close $P(u) = 0$ is to the exact steady state solution provided by $\Phi$, we proceeded as follows with $n = 4$. We chose $K^E_i$ and $K^F_i$ randomly from the uniform distribution on $[1,1000]\,\nM$ and $\log_{10}\lambda_i$ randomly from the uniform distribution on $[-3,3]$. We set $\et = \ft$ and chose $\log_{10}\et$ and $\log_{10}\st$ randomly from the uniform distribution on $[0,4]$, corresponding to a concentration range of $[1-10000]\,\nM$. We generated 10,000 such systems, for which we solved both $\Phi$ and $P(u)$ for the steady states. We found 108 systems for which the number of steady states differed between $\Phi$ and $P(u)$. We first set those aside but analyse them further below. For the remaining systems, we calculated $[E]/[F]$ for each steady state coming from $\Phi$ and listed them in order of increasing $[E]/[F]$: $s_1 < s_2 < \cdots < s_k$, where $k$ is the number of steady states. (We found $k = 1$ and $k = 3$ only, with no $k = 5$.) We matched these with the ordered list of positive solutions of $P(u) = 0$, $a_1 < a_2 < \cdots < a_k$. We measured the discrepancy between the exact solution coming from $\Phi$ and the approximate solution coming from $P(u)$ by calculating the average normalised difference,
\begin{equation}
\sigma = \frac{1}{k}\sum_{i=1}^k \frac{|s_i - a_i|}{s_i} \,.
\label{e-sig}
\end{equation}
Figure~\ref{f-apx}A shows that for nearly 80\% of the randomly chosen systems, the approximation is good to within $\sigma < 0.1$, irrespective of the values of $\st$ and $\et$. Figure~\ref{f-apx}B shows that the approximation gets steadily better as $\st/\et$ increases from $1$. We took $\st/\et \geq 5$ as our cut-off. In this range, $\sigma < 0.23$.

We then considered the 108 omitted systems for which $\Phi$ and $P(u)$ differed in the number of roots found. A histogram of these is plotted against $\log_{10} \st/\et$ on the bottom of Figure~\ref{f-apx}B. We found 52 miscounted systems for which $\st/\et \geq 5$. We examined each of these by hand and determined, on a conservative basis, that 45 of them were caused by numerical errors in $\Phi$. That is, when these systems were re-computed with finer tolerances and a denser set of initial conditions, the number of steady states was found to converge and to agree with those obtained from $P(u)$. The remaining 7 systems were adjudged to be possible errors arising from using $P(u)$ as an approximation for $\Phi$. Since there were 3385 systems for which $\st/\et \geq 5$, this gives a miscounting rate for $P(u)$ of $0.2\%$.

\subsubsection*{Any polynomial can be $P(u)$}

Suppose given any polynomial
\begin{equation}
Q(u) = A_{n+1}u^{n+1} + A_n u^n + \cdots + A_1 u + A_0
\label{e-poly2}
\end{equation}
with real coefficients such that $A_{n+1} > 0$ and $A_0 < 0$. We claim that for appropriate choice of $K^E_i$, $K^F_i$, $\lambda_i$ and $w = \et/\ft$, as well as $\st$ chosen arbitrarily, all positive, the corresponding $P(u)$ polynomial defined by (\ref{e-coef}) coincides with $Q(u)$. We show this by induction.

Note first that the term in square brackets in (\ref{e-coef}) can be rewritten as
\[ \left(1 + \frac{\st}{K^E_i}\right) - \lambda_i w\left(1 + \frac{\st}{K^F_{i+1}}\right) \,.\]
Start by choosing $\st > 0$ arbitrarily. Choose $w = -A_0 > 0$. For $0 \leq i \leq n-2$, choose $K^E_i$, $K^F_{i+1}$ and $\lambda_i$ inductively so that
\[ \left(1 + \frac{\st}{K^E_i}\right) - \lambda_i w\left(1 + \frac{\st}{K^F_{i+1}}\right) = \frac{A_{i+1}}{\lambda_0\cdots\lambda_{i-1}} = B_{i+1} \]
as follows. (When $i = 0$, the induction starts with $A_1 = B_1$ but the argument below is identical.) If $B_{i+1} = 0$, take $\lambda_i = 1/w$ and choose $K^E_i = K^F_{i+1} > 0$ arbitrarily. If $B_{i+1} > 0$, choose $K^E_i > 0$ so that 
\[ \left(1 + \frac{\st}{K^E_i}\right) > B_{i+1} \,,\]
which may always be done. Now choose $K^F_{i+1}$ and $\lambda_i$ so that
\begin{equation}
\left(1 + \frac{\st}{K^E_i}\right) - B_{i+1} = \lambda_i w\left(1 + \frac{\st}{K^F_{i+1}}\right) \,,
\label{e-b}
\end{equation}
which may also always be done. If $B_{i+1} < 0$ then $K^E_{i+1}$ may be chosen arbitrarily and the left hand side of (\ref{e-b} will always be positive. Hence, $K^F_{i+1}$ and $\lambda_i$ can always be chosen positive so that (\ref{e-b}) is satisfied. 

By following this inductive procedure for $0 \leq i \leq n-2$ we have chosen $\st$, $w$, $K^E_i$ for $0 \leq i \leq n-2$, $K^F_i$ for $1 \leq i \leq n-1$ and $\lambda_i$ for $0 \leq i \leq n-2$ all positive. With these choices we have satisfied (\ref{e-coef}) for all coefficients $A_i$ such that $0 \leq i < n$. Now consider the last two coefficients $A_n$ and $A_{n+1}$. Choose $\lambda_{n-1} = A_{n+1}/(\lambda_0\cdots\lambda_{n-2}) > 0$, so that (\ref{e-coef}) is satisfied for $A_{n+1}$. Now choose $K^E_{n-1}$ and $K^F_n$ such that 
\[ \frac{1}{K^E_{n-1}} - \frac{\lambda_{n-1}w}{K^F_n} = \frac{1}{\st}\left(\frac{A_n}{\lambda_0\cdots\lambda_{n-2}} - (1 - \lambda_{n-1}w)\right) = c \,,\]
as follows. The right hand side consists of terms like $A_n$, which are given, or terms that have been previously determined. Let $\alpha = \lambda_{n-1}w > 0$. We have to find $x, y > 0$ such that 
\[ x - \alpha y = c\,. \]
Since $\alpha > 0$, this can always be done for any $c$, thereby satisfying (\ref{e-coef}) for $A_n$. This completes the induction.

\subsubsection*{Numerical solution of $P(u) = 0$}

We used Matlab's {\tt roots} function, which is extremely fast and accurate. For $n$ up to 12 sites, $\sim 6000$ polynomials per second can be solved, giving a substantial improvement over numerical solution of $\Phi$.

\subsubsection*{Intractability of conditions for positive roots of $P(u)$}

Sturm's Theorem \citep{smiley} provides an algorithm for calculating the number of positive roots of a polynomial. We implemented this in the following Mathematica code (Wolfram Research, Champaign, IL, USA):
\begin{eqnarray*}
f_0[u\_] & := & \sum_{i = 0}^n a_i u^i \\
f_1[u\_] & := & \partial_u f_0[u] \\
f_k[u\_] & := & -\mbox{{\tt PolynomialRemainder}}[f_{k - 2}[u], f_{k - 1}[u], u] \,.
\end{eqnarray*}
Since the degree reduces by one with each remainder, the $f_k[u]$ must become constant for some $k \leq n$. Let $v(x)$ be the number of sign changes in the list $f_0[x], \cdots, f_n[x]$. Sturm's Theorem states that if $a < b$ are not roots of $f_0$ then the number of distinct roots of $f_0$ in $[a,b]$ equals $v(a) - v(b)$. We applied this to the general polynomial $a_3u^3 + a_2 u^2 + a_1 u + a_0$, corresponding to the case of just two sites, where we assumed that $a_3 > 0$ and $a_0 < 0$ in accordance with (\ref{e-coef}). We took the range to be $[0, \infty)$, using the fact that, for sufficiently large $b$, $v(b)$ is determined by the leading coefficients of $f_0, \cdots, f_n$. We found that the general polynomial has 3 positive roots if, and only if, the following conditions collectively hold:
\[ a_1 > 0, \hspace{1.5em}
-a_0 + \frac{a_1 a_2}{9 a_3} < 0, \hspace{1.5em}
\frac{2}{9}\left(-3 a_1 + \frac{a_2^2}{a_3}\right) > 0 \]
\[ \frac{9 a_3(a_1^2 a_2^2 - 4 a_1^3 a_3 + 18 a_0 a_1 a_2 a_3 - a_0(4 a_2^3 + 27 a_0 a_3^2))}{4(a_2^2 - 3 a_1 a_3)^2} > 0 \]
These show that the region in the space of coefficients which gives rise to the maximum number of positive roots is highly complex. The complexity increases extremely rapidly with $n$. For $n = 4$ the conditions are so unwieldy that even Mathematica cannot easily compute them. We concluded that the question of when multistability occurs for a given set of rate constants is mathematically intractable.

\subsubsection*{Model simulation}

We used the {\tt little b} computational infrastructure (Mallavarapu, Thomson, Ullian \& Gunawardena, submitted, 2007) to generate differential equation models. {\tt Little b} is a modular programming language in which models can be specified at a biological level of description and compiled into Matlab code, which can then be simulated. The system in Figure~\ref{f-rs}B was described in a {\tt little b} program, which was then instantiated for the required number of sites, making it unnecessary to write new Matlab code for different values of $n$. {\tt Little b} is freely available as open source software from {\sf littleb.org} and {\sf vcp.med.harvard.edu}. For simulations we used Matlab's {\tt ode15s} solver with absolute tolerance of $10^{-35}$. 

\normalsize

\section*{Acknowledgements}

We thank Rebecca Ward and Brian Seed for their comments on the manuscript; Aneil Mallavarapu for developing {\tt little b}; and the Department of Systems Biology for its support.

\bibliography{/home/jhcg/work/BIB/bio}

\begin{thebibliography}{10}

\bibitem{ag05}
G.~Altan-Bonnet and R.~N. Germain.
\newblock Modeling {T} cell antigen discrimination based on feedback control of
  digital {ERK} responses.
\newblock {\em PLoS Biol.}, 3:1925--38, 2005.

\bibitem{ajs98}
B.~Anderson, J.~Jackson, and M.~Sitharam.
\newblock Descartes' {R}ule of {S}igns revisited.
\newblock {\em Amer. Math. Monthly}, 105:447--451, 1998.

\bibitem{bf01}
C.~P. Bagowski and J.~E. Ferrell.
\newblock Bistability in the {JNK} cascade.
\newblock {\em Curr. Biol.}, 11(15):1176--82, 2001.

\bibitem{brs86}
A.~T. Bharucha-Reid and M.~Sambandham.
\newblock {\em Random Polynomials}.
\newblock Probability and Mathematical Statistics. Academic Press, Burlington,
  MA, USA, 1986.

\bibitem{bray95}
D.~Bray.
\newblock Protein molecules as computational elements in living cells.
\newblock {\em Nature}, 376:307--12, 1995.

\bibitem{bs97}
W.~R. Burack and T.~W. Sturgill.
\newblock The activating dual phosphorylation of {MAPK} by {MEK} is
  nonprocessive.
\newblock {\em Biochemistry}, 36:5929--33, 1997.

\bibitem{coh00}
P.~Cohen.
\newblock The regulation of protein function by multisite phosphorylation - a
  25 year update.
\newblock {\em Trends Biochem. Sci.}, 25:596--601, 2000.

\bibitem{cb95}
A.~Cornish-Bowden.
\newblock {\em Fundamentals of Enzyme Kinetics}.
\newblock Portland Press, London, UK, 2nd edition, 1995.

\bibitem{crick84}
F.~Crick.
\newblock Memory and molecular turnover.
\newblock {\em Nature}, 312:101, 1984.

\bibitem{ks98}
P.~de~Konnick and H.~Schulman.
\newblock Sensitivity of {CaM} {K}inase {II} to the frequency of
  $\mbox{Ca}^{2+}$ oscillations.
\newblock {\em Science}, 279(5):227--300, 1998.

\bibitem{dws07}
D.~A. Drubin, J.~C. Way, and P.~A. Silver.
\newblock Designing biological systems.
\newblock {\em Genes Dev.}, 21:242--54, 2007.

\bibitem{ek95}
A.~Edelman and E.~Kostlan.
\newblock How many zeros of a random polynomial are real?
\newblock {\em Bull. Amer. Math. Soc.}, 32:1--37, 1995.

\bibitem{fer02}
J.~E. Ferrell.
\newblock Self-perpetuating states in signal transduction: positive feedback,
  double-negative feedback and bistability.
\newblock {\em Current Opinion in Chemical Biology}, 6:140--8, 2002.

\bibitem{fb97}
J.~E. Ferrell and R.~R. Bhatt.
\newblock Mechanistic studies of the dual phosphorylation of mitogen-activated
  protein kinase.
\newblock {\em J. Biol. Chem.}, 272:19008--16, 1997.

\bibitem{fm98}
J.~E. Ferrell and E.~M. Machleder.
\newblock The biochemical basis of an all-or-none cell fate switch in {{\em
  Xenopus}} oocytes.
\newblock {\em Science}, 280:895--8, 1998.

\bibitem{flsa06}
C.~M. Furdui, E.~D. Lew, J.~Schlessinger, and K.~S. Anderson.
\newblock Autophosphorylation of {FGFR1} kinase is mediated by a sequential and
  precisely ordered reaction.
\newblock {\em Molecular Cell}, 21:711--17, 2006.

\bibitem{gsd04}
S.~Graziani, P.~Silar, and M.-J. Daboussi.
\newblock Bistability and hysteresis of the '{S}ecteur' differentiation are
  controlled by a two-gene locus in {{\em Nectria haematocca}}.
\newblock {\em BMC Biology}, 2, 2004.
\newblock doi:10.1186/1741-7007-2-18.

\bibitem{gun05}
J.~Gunawardena.
\newblock Multisite protein phosphorylation makes a good threshold but can be a
  poor switch.
\newblock {\em Proc. Natl. Acad. Sci. USA}, 102:14617--22, 2005.

\bibitem{har01}
A.~J. Harwood.
\newblock Regulation of {GSK-3}: a cellular multiprocessor.
\newblock {\em Cell}, 105:821--4, 2001.

\bibitem{hs74}
M.~W. Hirsch and S.~Smale.
\newblock {\em Differential Equations, Dynamical Systems and Linear Algebra}.
\newblock Pure and Applied Mathematics. Academic Press, San Diego, USA, 1974.

\bibitem{htes02}
C.~I. Holmberg, S.~E.~F. Tran, J.~E. Eriksson, and L.~Sistonen.
\newblock Multisite phosphorylation provides sophisticated regulation of
  transcription factors.
\newblock {\em Trends Biochem. Sci.}, 27:619--27, 2002.

\bibitem{hmu06}
J.~E. Hopcroft, R.~Motwani, and J.~D. Ullman.
\newblock {\em Introduction to Automata Theory, Languages, and Computation}.
\newblock Addison Wesley, Boston, MA, USA, 2006.

\bibitem{hf96}
C.-Y.~F. Huang and J.~E. Ferrell.
\newblock Ultrasensitivity in the mitogen-activated protein kinase cascade.
\newblock {\em Proc. Natl. Acad. Sci. USA}, 93:10078--83, 1996.

\bibitem{jceoh05}
R.~J. Johnston, S.~Chang, J.~F. Etchberger, C.~O. Ortiz, and O.~Hobert.
\newblock Micro{RNA}s acting in a double-negative feedback loop to control a
  neuronal cell fate decision.
\newblock {\em Proc. Natl. Acad. Sci. USA}, 102:12449--54, 2005.

\bibitem{kan03}
E.~Kandel.
\newblock The molecular biology of memory storage: a dialog between genes and
  synapses.
\newblock In H.~J{\"o}rnvall, editor, {\em Nobel Lectures, Physiology and
  Medicine 1996-2000}. World Scientific, 2003.

\bibitem{ken03}
P.~J. Kennelly.
\newblock Archaeal protein kinases and phosphatases: insights from genomics and
  biochemistry.
\newblock {\em Biochem. J.}, 370:373--89, 2003.

\bibitem{kc96}
D.~M. Kim and C.~Y. Choi.
\newblock A semicontinuous prokaryotic coupled transcription/translation system
  using a dialysis membrane.
\newblock {\em Biotechnol Prog.}, 12:645--9, 1996.

\bibitem{kf07}
S.~Y. Kim and J.~E. Ferrell.
\newblock Substrate competition as a source of ultrasensitivity in the
  inactivation of {W}ee1.
\newblock {\em Cell}, 128:1133--45, 2007.

\bibitem{kpt03}
P.~Klein, T.~Pawson, and M.~Tyers.
\newblock Mathematical modeling suggests cooperative interactions between a
  disordered polyvalent ligand and a single receptor site.
\newblock {\em Current Biology}, 13:1669--78, 2003.

\bibitem{kz07}
R.~J. Klose and Y.~Zhang.
\newblock Regulation of histone methylation by demethylimination and
  demethylation.
\newblock {\em Nat. Rev. Mol. Cell Biol.}, 8:307--18, 2007.

\bibitem{lrsglea}
G.~Lahav, N.~Rosenfeld, A.~Sigal, N.~Geva-Zatorsky, A.~J. Levine, M.~B.
  Elowitz, and U.~Alon.
\newblock Dynamics of the p53-mdm2 feedback loop in individual cells.
\newblock {\em Nature Genetics}, 36:147--50, 2004.

\bibitem{lbs00}
A.~Levchenko, J.~Bruck, and P.~W. Sternberg.
\newblock Scaffold proteins may biphasically affect the levels of
  mitogen-activated protein kinase signaling and reduce its threshold
  properties.
\newblock {\em Proc. Natl. Acad. Sci.}, 97:5818--23, 2000.

\bibitem{lls03}
J.~Lisman, J.~W. Lichtman, and J.~R. Sanes.
\newblock {LTP}: perils and progress.
\newblock {\em Nature Rev. Neuro.}, 4:926--9, 2003.

\bibitem{lr06}
J.~Lisman and S.~Raghavachari.
\newblock A unified model of the presynaptic and postsynaptic changes during
  {LTP} at {CA1} synapses.
\newblock {\em Sci. STKE}, 356, 2006.
\newblock doi:10.1126/stke.3562006re11.

\bibitem{lisman85}
J.~E. Lisman.
\newblock A mechanism for memory storage insensitive to molecular turnover: a
  bistable autophosphorylating kinase.
\newblock {\em Proc. Natl. Acad. Sci. USA}, 82:3055--7, 1985.

\bibitem{lsk02}
J.~E. Lisman, H.~Schulman, and H.~Kline.
\newblock The molecular basis of {CaMKII} function in synaptic and behavioural
  memory.
\newblock {\em Nat. Rev. Neurosci.}, 3:175--90, 2002.

\bibitem{mmogkjm}
B.~Macek, I.~Mijakovic, J.~V. Olsen, F.~Gnad, C.~Kumar, P.~R. Jensen, and
  M.~Mann.
\newblock The serine/threonine/tyrosine phosphoproteome of the model bacterium
  {\em {b}acillus subtilis}.
\newblock {\em Mol. Cell. Proteomics}, 2007.
\newblock doi:10.1074/mcp.M600464-MCP200.

\bibitem{mogsjp}
M.~Mann, S.-E. Ong, M.~Gr{\o}nberg, H.~Steen, O.~N. Jensen, and A.~Pandey.
\newblock Analysis of protein phosphorylation using mass spectrometry:
  deciphering the phosphoproteome.
\newblock {\em Trends Biotechnol.}, 20:261--8, 2002.

\bibitem{mhk04}
N.~I. Markevich, J.~B. Hoek, and B.~N. Kholodenko.
\newblock Signalling switches and bistability arising from multisite
  phosphorylation in protein kinase cascades.
\newblock {\em J. Cell Biol.}, 164:353--9, 2004.

\bibitem{mq07}
J.~Melin and S.~R. Quake.
\newblock Microfluidic large-scale integration: the evolution of design rules
  for biological large-scale integration.
\newblock {\em Annu. Rev. Biophys. Biomol. Struct.}, 36:213--31, 2007.

\bibitem{mj61}
J.~Monod and F.~Jacob.
\newblock General conclusions: teleonomic mechanisms in cellular metabolism,
  growth and differentiation.
\newblock {\em Cold Spring Harbor Symp. Quant. Biol.}, 26:389--401, 1961.

\bibitem{niinmiok}
M.~Nakajima, K.~Imai, H.~Ito, T.~Nishiwaki, Y.~Murayama, H.~Iwasaki, T.~Oyama,
  and T.~Kondo.
\newblock Reconstitution of circadian oscillation of cyanobacterial {KaiC}
  phosphorylation in vitro.
\newblock {\em Science}, 308:414--5, 2005.

\bibitem{ogmkc}
F.~Ortega, J.~L. Garc\'{e}s, F.~Mas, B.~N. Kholodenko, and M.~Cascante.
\newblock Bistability from double phosphorylation in signal transduction.
\newblock {\em FEBS J.}, 273:3915--26, 2006.

\bibitem{oud-ea04}
E.~M. Ozbudak, M.~Thattai, H.~N. Lim, B.~I. Shraiman, and A.~van Oudenaarden.
\newblock Multistability in the lactose utilization network of {{\em
  Escherichia coli}}.
\newblock {\em Nature}, 427:737--40, 2004.

\bibitem{rrk06}
M.~Rape, S.~K. Reddy, and M.~W. Kirschner.
\newblock The processivity of multiubiquitination by the {APC} determines the
  order of substrate degradation.
\newblock {\em Cell}, 124:89--103, 2006.

\bibitem{rtvrb}
A.~R. Reynolds, C.~Tischer, P.~J. Verveer, O.~Rocks, and P.~I.~H. Bastiaens.
\newblock {EGFR} activation coupled to inhibition of tyrosine phosphatases
  causes lateral signal propagation.
\newblock {\em Nature Cell Biol.}, 5:447--53, 2003.

\bibitem{rsi05}
D.~M. Rothman, M.~D. Shults, and B.~Imperiali.
\newblock Chemical approaches for investigating phosphorylation in signal
  transduction networks.
\newblock {\em Trends Cell Biol.}, 15:502--10, 2005.

\bibitem{sh03}
C.~Salazar and T.~H{\"o}fer.
\newblock Allosteric regulation of the transcription factor {NFAT1} by multiple
  phosphorylation sites: a mathematical analysis.
\newblock {\em J. Mol. Biol.}, 327:31--45, 2003.

\bibitem{sh07}
C.~Salazar and T.~H{\"o}fer.
\newblock Versatile regulation of multisite protein phosphorylation by the
  order of phosphate processing and protein-protein interactions.
\newblock {\em FEBS J.}, 274:1046--60, 2007.

\bibitem{smiley}
M.~F. Smiley.
\newblock A proof of {S}turm's {T}heorem.
\newblock {\em Amer. Math. Monthly}, 49:185--6, 1942.

\bibitem{sbroa}
A.~S. Spirin, V.~I. Baranov, L.~A. Ryabova, S.~Y. Ovodov, and Y.~B. Alakhov.
\newblock A continuous cell-free translation system capable of producing
  polypeptides in high yield.
\newblock {\em Science}, 242:1162--4, 1988.

\bibitem{sjrmk}
H.~Steen, J.~A. Jebanathirajah, J.~Rush, N.~Morrice, and M.~W. Kirschner.
\newblock Phosphorylation analysis by mass spectrometry: myths, facts, and the
  consequences for qualitative and quantitative measurements.
\newblock {\em Mol. Cell Proteomics}, 5:172--81, 2005.

\bibitem{srg00}
A.~M. Stock, V.~L. Robinson, and P.~N. Goudreau.
\newblock Two component signal transduction.
\newblock {\em Annu. Rev. Biochem.}, 69:183--215, 2000.

\bibitem{um06}
R.~Unger and J.~Moult.
\newblock Towards computing with proteins.
\newblock {\em Proteins}, 63:53--64, 2006.

\bibitem{vwrb00}
P.~J. Verveer, F.~S. Wouters, A.~R. Reynolds, and P.~I. Bastiaens.
\newblock Quantitative imaging of lateral {ErbB1} receptor signal propagation
  in the plasma membrane.
\newblock {\em Science}, 290:1567--70, 2000.

\bibitem{wsX704}
L.~Wang and E.~Sontag.
\newblock A remark on the number of steady states in a multiple futile cycle.
\newblock arXiv:0704.0036v1, 2007.

\bibitem{xf03}
W.~Xiong and J.~E. Ferrell.
\newblock A positive-feedback-based bistable 'memory module' that governs a
  cell fate decision.
\newblock {\em Nature}, 426:460--5, 2003.

\bibitem{zz01}
Y.~Zhao and Z.-Y. Zhang.
\newblock The mechanism of dephosphorylation of extracellular signal-regulated
  kinase 2 by mitogen-activated protein kinsae phosphatase 3.
\newblock {\em J. Biol. Chem.}, 276:32382--91, 2001.

\end{thebibliography}
\bibliographystyle{plain}

\newpage

\newpage

\newpage

\newpage

\newpage

\newpage

\end{document}